\documentclass[conference]{IEEEtran}
\IEEEoverridecommandlockouts
\usepackage{cite}
\usepackage{amsmath,amssymb,amsfonts}
\usepackage{algorithmic}
\usepackage{graphicx}
\usepackage{textcomp}
\usepackage{xcolor}
\usepackage{listings}    
\usepackage{enumitem}
\usepackage{booktabs}
\usepackage{accessibility}
\usepackage[hidelinks]{hyperref}
\usepackage[nameinlink]{cleveref}
\usepackage{pifont}%
\usepackage{tcolorbox}
\usepackage{threeparttable}
\usepackage{xcolor}
\usepackage[utf8]{inputenc}
\usepackage{float}
\usepackage{authblk}

\usepackage{framed}

\def\BibTeX{{\rm B\kern-.05em{\sc i\kern-.025em b}\kern-.08em
    T\kern-.1667em\lower.7ex\hbox{E}\kern-.125emX}}
\begin{document}

\title{%
Separating Secrets from Placeholders: A Hybrid CNN-CodeBERT Framework for Three-Class Credential Leakage Detection\\
}

\author[1]{Maksuda Bilkis Baby}
\author[2]{Khusika Shah}
\author[1]{Naiyue Liang}  %
\author[1]{Lei Zhang}

\affil[1]{Information Systems, University of Maryland, Baltimore County, USA}
\affil[2]{Computer Science and Electrical Engineering, University of Maryland, Baltimore County, USA}
\affil[ ]{\textit{mstmakb1, kshah7, vs76219, leizhang@umbc.edu}}

\maketitle
\begin{abstract}
Credential leakage in public source code repositories poses a critical security threat, with over 23.8 million secrets exposed in 2024 alone. Existing detection tools suffer from high false-positive rates because rigid pattern matching and binary classification schemes fail to distinguish genuine credentials from placeholder or weak credentials. We propose a three-class classification framework that explicitly models placeholder or weak credentials as a distinct class, leveraging CodeBERT-based semantic understanding combined with character-level pattern recognition. We evaluate our approach on a newly constructed dataset of 9,426 samples spanning 10 programming languages. Our model achieves a Matthews Correlation Coefficient of 0.86 and a macro F1-score of 0.90, achieving 93\% recall and 89\% precision for genuine credential leaks while reducing high-severity alerts by 33.0\% (from 373 to 250) without sacrificing security coverage. Compared to prior character-level approaches, our method improves placeholder or weak credential detection from 54\% to 81\% F1-score while maintaining strong cross-language generalization, with 9 of 10 languages achieving F1 above 0.80 under leave-one-language-out evaluation.
\end{abstract}


\section{Introduction}
\label{sec:intro}

Public repositories have become an integral part of modern software development. Platforms such as GitHub enable large-scale collaboration, continuous integration, and rapid iteration. As projects evolve through commits, forks, and dependencies, maintaining security becomes increasingly difficult. When repositories become public, one persistent problem is credential leakage, in which passwords, API keys, and secret tokens are embedded directly in source code~\cite{feng2022icse,rahman2026issueguard}, creating opportunities for attackers to scan and harvest them.

The scale of this issue continues to grow. The GitGuardian 2025 State of Secrets Sprawl Report states that 23.8 million new secrets were exposed in public GitHub repositories during 2024, a 25\% increase from the previous year~\cite{gitguardian2025}. The report estimates that remediation requires two person-hours per incident, resulting in approximately \$5.7 billion in global labor cost~\cite{gitguardian2025}. Exposed credentials often persist across commits and propagate through forks, increasing long-term maintenance effort. Real-world breaches illustrate the impact. In the 2016 Uber incident, attackers used exposed GitHub credentials to access hardcoded Amazon Web Services (AWS) keys, ultimately compromising internal systems and exposing data from 57 million users~\cite{uber2016breach}.

In the literature, empirical studies confirm that credential leakage is widespread and persistent. Meli et al.\ report that hardcoded secrets frequently remain exposed long after disclosure~\cite{meli2019ndss}. Feng et al.\ show that over 82\% of leaked passwords remain accessible for at least 16 days~\cite{feng2022icse}. These findings highlight the urgent need for automated detection techniques that scale across diverse languages while remaining precise enough for real-world maintenance workflows.

Existing credential leakage detection (CLD) approaches fall into two broad categories. First, rule-based tools, such as Gitleaks~\cite{gitleaks} and TruffleHog~\cite{trufflehog}, rely on regular expressions and entropy heuristics. Although efficient, they often suffer from low recall or excessive false positives~\cite{basak2023esem}. Basak et al.\ report that even the best-performing configuration achieves 75\% precision with only 36\% recall~\cite{basak2023esem}. Second, learning-based approaches attempt to improve robustness. PassFinder uses character-level convolutional neural networks (CNNs) trained on password corpora and code context and reports 81.54\% precision and 80.51\% recall across ten languages~\cite{feng2022icse}. Large language models (LLMs) have also been explored for security analysis~\cite{pearce2022asleep, sandoval2023lost}. However, their black-box behavior and deployment cost limit practical integration into continuous integration pipelines.

Existing methods for CLD face three major limitations in maintenance settings.

First, most approaches treat credential detection as a \emph{binary task}. They force ambiguous cases into either genuine leaks or benign code. Placeholder values such as \texttt{API\_KEY = `your\_api\_key\_here'} and weak test credentials such as \texttt{password = 12345} with \texttt{user = `user1'} resemble real secrets but often pose minimal risk. Binary framing increases false positives, creates alert fatigue, and reduces developer trust~\cite{chess2004ieee, rahman2022emse}.

Second, many models rely on surface patterns without \emph{reasoning/semantic understanding} about how strings function in context. For example, \texttt{secret = `your-secret'} may appear as a default configuration value in an Express.js session setup. A syntactic detector flags it as a leak because it associates a hardcoded string with a security field. In practice, it may be a placeholder. Effective detection requires understanding how the string interacts with the surrounding code and what security boundary it affects~\cite{meli2019ndss, basak2023esem}.

Lastly, detection \emph{performance varies across programming languages}. Differences in syntax, naming conventions, and string literals reduce generalization. Tools tuned for one language often degrade on others~\cite{feng2022icse, allamanis2018survey}. As projects span multiple ecosystems, consistent cross-language performance becomes essential.

We address these limitations through a three-class formulation that models ambiguity explicitly and incorporates semantic reasoning over source code. Our goal is to improve triage precision while maintaining strong coverage of genuine leaks across languages. Our main \textbf{contributions} are as follows.
\begin{enumerate}[leftmargin=*]
\item We introduce a three-class formulation that distinguishes (i) \textit{No Leak Credentials}, (ii) \textit{Genuine Credentials}, and (iii) \textit{Placeholder (or Weak Credentials)}, reducing false positives while preserving recall.
\item We construct and release a manually annotated dataset of 9,426 snippets across ten programming languages with high inter-annotator agreement (Cohen's $\kappa$ = 0.89)
 on Zenodo (\href{https://doi.org/10.5281/zenodo.18881159}{doi:10.5281/zenodo.18881159}) to ensure reproducibility. Note that sensitive information has been sanitized to prevent exposure of credentials.
\item We demonstrate that semantic representations from CodeBERT enable effective classification across all three classes, including placeholder disambiguation and genuine credential leak detection. Character-level pattern modeling provides complementary coverage, and a lightweight language adapter improves performance across 8 of 10 languages.
\end{enumerate}

We aim to address the following research questions (\textbf{RQs}):

\noindent\textbf{RQ1: Dataset Quality and Reliability.}
To what extent can a manually annotated dataset reliably capture credential occurrences across multiple programming languages?

\noindent\textbf{RQ2: Three-Class Effectiveness.}
How effectively does the proposed three-class formulation detect genuine credential leaks while distinguishing them from placeholder or weak credentials?

\noindent\textbf{RQ3: Cross-Language Generalization.}
How well does the model generalize across programming languages, and what role does the language adapter play?

\noindent\textbf{RQ4: Architectural Component Contributions.}
What is the contribution of each architectural component to overall detection performance?

\noindent\textbf{RQ5: Comparison with State-of-the-Art Systems.}
How does the approach compare with state-of-the-art CLD systems under a unified three-class
evaluation?

The remainder of the paper is organized as follows. Section~\ref{sec:related-work} reviews related work. Section~\ref{sec:dataset} describes dataset construction. Section~\ref{sec:method} presents the proposed approach. Section~\ref{sec:experiments} details the experimental setup. Section~\ref{sec:results} reports results. Section~\ref{sec:threats} discusses threats to validity, and Section~\ref{sec:conclusion} concludes the paper.

\section{Related Work}\label{sec:related-work}
This section reviews previous work on empirical measurement, detection techniques, and foundational approaches to semantic code understanding.

\subsection{Empirical Studies and Measurement}

Large-scale studies show that hardcoded credentials frequently appear in public repositories and often remain valid long enough to pose serious risks. Meli et al.\ demonstrated that exposed secrets persist over time in GitHub repositories, creating opportunities for compromise~\cite{meli2019ndss}, and related work confirms that security-sensitive hardcoded values remain prevalent in practice~\cite{rahman2019seven}. Although these studies establish the severity of secret leakage, they focus on ecosystem-level measurement and do not provide labeled datasets for training or evaluation.

Subsequent research shows that leaked credentials propagate through forks and dependencies, while attackers actively monitor repositories for new leaks~\cite{zhou2025sp,zhang2023don,shi2025skeleton}. Leakage extends beyond source code to container images and mobile applications~\cite{dahlmanns2023secrets,schmidt2025leaky}, and insecure infrastructure-as-code patterns recur across projects~\cite{rahman2019seven}. Despite broadening the understanding of credential exposure, these works largely adopt a binary notion of leakage and do not address ambiguity from placeholders.

\subsection{Detection Systems and Techniques}
Most credential detection tools treat the task as binary classification, labeling credential-like strings as genuine leaks or benign code~\cite{
meli2019ndss,
feng2022icse,
gitleaks,
trufflehog,
basak2023esem,
pearce2022asleep,
sandoval2023lost,
chess2004ieee,
rahman2022emse,
zhou2025sp,
zhang2023don,
shi2025skeleton,
dahlmanns2023secrets,
schmidt2025leaky,
saha2020comsnets
}. However, real-world code includes ambiguous cases such as placeholders (e.g., \texttt{YOUR\_API\_KEY}) and weak credentials (e.g., \texttt{pwd = `abcd' with user = `dummy\_user'}) that resemble real secrets but pose no actionable risk. Misclassification increases false positives and warning overload. Prior studies show that excessive false positives and poor prioritization reduce developer trust and long-term adoption~\cite{johnson2013don,ayewah2007evaluating,ayewah2008using,ayewah2010google,rahman2022secret}. These findings underscore the need to distinguish ambiguous credential-like strings while preserving actionable warnings.

Existing secret detection techniques are categorized into rule-based and learning-based approaches~\cite{basak2023esem}. Rule-based methods rely on predefined patterns, whereas learning-based methods infer credential characteristics from data. We will discuss them in detail in the following paragraphs.  %

Rule-based secret detection systems rely on manually crafted patterns, regular expressions, and entropy-based heuristics to identify credential-like strings in source code~\cite{gitleaks,trufflehog}. Although widely adopted, these rule-based approaches lack contextual reasoning, leading to precision below 75\% in large-scale evaluations due to difficulty distinguishing placeholders from genuine credentials~\cite{basak2023esem}. This limitation produces numerous false positives and low-value alerts, contributing to warning fatigue and reduced practical effectiveness.

Learning-based methods improve credential detection by modeling credential patterns from data through neural representations or contextual reasoning~\cite{saha2020comsnets}, such as PassFinder’s character-level representations~\cite{feng2022icse} and secret–asset relationship modeling for contextual triage~\cite{basak2024assetharvester}. Although these methods outperform rule-based systems, they remain binary and do not model dataset or label-level ambiguity. Recent work also applies LLMs such as ChatGPT to security analysis~\cite{zhou2024llmvuln}, but deployment challenges---including high cost, latency, limited transparency~\cite{bommasani2021foundation}, confidentiality concerns, and risks of insecure or over-trusted outputs~\cite{pearce2022asleep,sandoval2023lost}---limit their practicality, motivating specialized models that better balance accuracy, efficiency, and interpretability.

A major barrier to advancing detection techniques is the lack of publicly available labeled datasets. Most measurement studies do not release annotated data for supervised learning~\cite{meli2019ndss,zhou2025sp}, and PassFinder did not publish its full annotated dataset due to security and privacy concerns~\cite{feng2022icse}. Existing benchmarks such as CodeSearchNet and Vault support semantic code understanding but do not specifically address credential detection or distinguish genuine credential leaks from placeholders and benign code~\cite{husain2019codesearchnet,nguyen2023vault}. This scarcity of annotated datasets limits the training, evaluation, and comparison of detection models under realistic conditions.

\subsection{Foundations for Semantic and Cross-Language Detection}

Semantic code representation models underpin credential detection, and surveys highlight their effectiveness for code intelligence tasks~\cite{allamanis2018survey}. CodeBERT introduced joint pretraining on code and natural language~\cite{feng2020codebert}, and GraphCodeBERT incorporated data-flow information to capture deeper semantics~\cite{guo2021graphcodebert}. Models such as PLBART and UniXcoder further support cross-language understanding~\cite{ahmad2021unified,guo2022unixcoder}. Although these approaches improve tasks like defect prediction and code search~\cite{allamanis2018survey}, they are not designed to distinguish ambiguous credential-like strings from genuine secrets.

Cross-language detection is challenging due to syntactic and convention differences. PassFinder showed that performance varies significantly across languages despite shared character-level features~\cite{feng2022icse}. Recent code representation models reduce such discrepancies through shared semantics and multilingual pretraining. CodeBERT~\cite{feng2020codebert}, GraphCodeBERT~\cite{guo2021graphcodebert}, and UniXcoder~\cite{guo2022unixcoder} improve cross-language generalization, but they do not explicitly address ambiguity in credential detection.

In contrast to prior work, our approach formulates CLD as a three-class classification problem that distinguishes genuine credentials, placeholders, and benign code. We construct and release a manually annotated dataset to support reproducible evaluation. Our hybrid method leverages snippet-level semantic context to distinguish genuine leaks from placeholders, reducing alert fatigue while maintaining strong detection coverage, with character-level pattern recognition jointly capturing surface-level string characteristics. We further evaluate cross-language generalization across ten programming languages, demonstrating consistent performance despite syntactic diversity. These contributions address key limitations of prior binary detection approaches and improve practical credential detection workflows.

\section{Dataset Construction and Annotation}
\label{sec:dataset}

\begin{table}[t]
\centering
\caption{Overall class distribution in the proposed dataset.}
\label{tab:class_distribution}
\begin{tabular}{lrr}
\toprule
\textbf{Class} & \textbf{Count} & \textbf{Proportion} \\
\midrule
No Leak(Class 0)       & 5{,}903 & 62.6\% \\
Genuine Leak (Class 1)        & 2{,}363 & 25.1\% \\
Placeholder(Class 2)  & 1{,}160 & 12.3\% \\
\midrule
\textbf{Total}                & 9{,}426 & 100\%  \\
\bottomrule
\end{tabular}
\end{table}

We build a manually annotated dataset to evaluate our CLD system. We collect credential-related code snippets from public GitHub repositories. The dataset supports the three-class formulation from Section~\ref{sec:intro}: (i) \textit{No Leak}, (ii) \textit{Genuine Credential Leak}, and (iii) \textit{Placeholder}. This section describes the data collection process, candidate extraction strategy, annotation schema, annotation workflow, and dataset statistics.

\begin{table*}[t]
\centering
\caption{Per-language and per-class sample distribution in the proposed dataset.}
\label{tab:language_class_distribution}
\begin{tabular}{lrrrrr}
\toprule
\textbf{Language} & \textbf{No Leak} & \textbf{Genuine Leak} & \textbf{Placeholder} & \textbf{Total} \\
\midrule
C          & 763 (86.8\%) &  58 (6.6\%)  &  58 (6.6\%)  &  879 \\
C\#        & 654 (69.2\%) & 218 (23.1\%) &  73 (7.7\%)  &  945 \\
C++        & 799 (83.7\%) &  95 (9.9\%)  &  61 (6.4\%)  &  955 \\
Go         & 838 (88.0\%) &  72 (7.6\%)  &  42 (4.4\%)  &  952 \\
Java       & 348 (36.7\%) & 385 (40.6\%) & 215 (22.7\%) &  948 \\
JavaScript & 332 (34.3\%) & 439 (45.4\%) & 196 (20.3\%) &  967 \\
PHP        & 640 (66.3\%) & 229 (23.7\%) &  97 (10.0\%) &  966 \\
Python     & 295 (30.7\%) & 503 (52.3\%) & 163 (17.0\%) &  961 \\
Ruby       & 547 (60.6\%) & 198 (22.0\%) & 157 (17.4\%) &  902 \\
TypeScript & 687 (72.2\%) & 166 (17.5\%) &  98 (10.3\%) &  951 \\
\midrule
\textbf{Total} & 5{,}903 (62.6\%) & 2{,}363 (25.1\%) & 1{,}160 (12.3\%) & 9{,}426 \\
\bottomrule
\end{tabular}
\end{table*}

\subsection{Data Collection and Candidate Extraction}
We collect candidate code snippets from public GitHub repositories using the GitHub REST API via the PyGithub client~\cite{pygithub}. Following prior work, we harvest candidates through GitHub's code search interface (rather than sampling repositories directly), as it scales to large corpora and better reflects how developers locate code in practice~\cite{kalliamvakou2014promises,cosentino2017systematic}.
Searches are scoped to 10 programming languages: C, C++, C\#, Go, Java, 
JavaScript, PHP, Python, Ruby, and TypeScript using the \texttt{language} 
qualifier. These languages span diverse syntax styles, type systems, and 
paradigms, allowing us to assess whether the three-class formulation 
generalizes across heterogeneous development 
contexts~\cite{allamanis2018survey, feng2022icse}.

Search queries combine credential-related seed terms (\texttt{password}, 
\texttt{token}, \texttt{secret}, \texttt{api\_key}) with contextual terms 
covering email domains, infrastructure identifiers (e.g., \texttt{host}, 
\texttt{database}), account-related terms (e.g., \texttt{login}, 
\texttt{authenticate}), and specific phrases such as \texttt{client\_secret} 
and \texttt{basic authentication}. Queries are randomly shuffled before 
execution to improve diversity. This keyword strategy follows prior detection 
work and applies only to candidate collection, not to training or 
evaluation~\cite{meli2019ndss, feng2022icse, basak2023esem}.

Snippets are deduplicated by hashing the context window. The collector 
targets up to 1,000 unique snippets per language, rotating across a pool 
of API tokens when the rate-limit quota falls below a safety threshold. 
Each snippet is stored with metadata including programming language, 
repository name, file path, GitHub URL, matched line, and collection 
timestamp; the \texttt{label} column is left empty for manual annotation, 
as described next.

\subsection{Annotation Schema}

Existing binary formulations fail to distinguish genuine credentials from 
placeholder values, producing large volumes of false alerts in 
practice~\cite{johnson2013don, ayewah2010google, basak2023esem, rahman2022secret}. 
We propose a three-class schema that explicitly models this ambiguity. Each 
candidate snippet is assigned one label based on its security significance 
and likelihood of enabling unauthorized access:

\begin{itemize}[leftmargin=*]
\item \textbf{No Leak (Class 0):} No literal credential value is present. 
Includes indirect references, environment-variable lookups, schema definitions, 
and external configuration retrieval. Typical examples: 
\texttt{api\_key = os.getenv("API\_KEY")}, 
\texttt{config["secret"] = get\_secret("DB\_PASSWORD")}, 
\texttt{class User: password: str}.

\item \textbf{Genuine Credential Leak (Class 1):} A hardcoded 
value that appears to represent a valid authentication secret 
within the code context. It covers human- or machine-generated 
passwords, API keys, secret tokens, and client secrets 
(e.g., \texttt{\detokenize{password="to*****08"}}, 
\texttt{\detokenize{api_key="AIzaSyD*****mE"}}).\footnote{In all examples in this paper, we mask sensitive substrings using \texttt{*} to prevent exposing real credentials and to protect security and privacy.} Weak passwords 
paired with realistic usernames or email addresses also fall 
into this class, as the combination carries real access risk 
even when the password itself seems weak 
(e.g., \texttt{\detokenize{user="m****@outlook.com"}}; 
\texttt{\detokenize{pwd="abc123"}}).

\item \textbf{Placeholder (Class 2):} The value 
resembles a credential syntactically but clearly serves as a placeholder, example, or test artifact. This includes explicit placeholder indicators  (e.g., \texttt{\detokenize{password="Your_Pwd_Here"}} with 
\texttt{\detokenize{user="dummy_user"}}), and trivial credential pairs 
used for testing (e.g., \texttt{\detokenize{password="12345"}} with 
\texttt{\detokenize{user="user1"}}). Because such values follow the same 
syntactic patterns as real credentials (e.g., 
\texttt{\detokenize{password=...}} or \texttt{\detokenize{token=...}}), 
they are often captured during candidate extraction but can be distinguished 
through contextual cues indicating documentation, examples, or test 
configurations rather than deployable authentication secrets.
\end{itemize}

Since intent depends on context, identical strings may receive different 
labels depending on whether they appear in production paths, test fixtures, 
or documentation, reflecting ambiguity common in public 
repositories~\cite{basak2023esem}.

\textbf{Annotation Decision Strategy:}
Special attention is given to weak passwords that, when paired with 
realistic usernames or emails (e.g., \mbox{\texttt{user="dan****@gmail.com"; 
pwd="abcd"}}), are treated as Class 1 due to their potential 
for unauthorized access if deployed in real systems. However, when 
the username appears synthetic or generic (e.g., \mbox{\texttt{user="dummy\_user1"; 
pwd="abcd"}}), the snippet is treated as Class 2, as the 
combination suggests a placeholder rather than a real credential.

\subsection{Annotation Workflow and Quality Control}

At least three co-authors annotate the dataset and cross-examine 
the annotations. Two primary annotators label 
each sample, and a third annotator independently reviews the assigned 
label. We measure inter-annotator agreement on a randomly sampled 
subset of 4,000 snippets from the full 9,426-sample dataset using 
Cohen's $\kappa$ and obtain a value of 0.89, indicating near-perfect 
agreement~\cite{artstein2008inter}. After agreement assessment, the 
third annotator reviews all disagreement cases and performs a final 
adjudication pass, resolving conflicts, enforcing consistent 
application of the labeling guidelines across programming languages, 
and correcting systematic inconsistencies (e.g., ensuring weak 
passwords are treated consistently when paired with realistic 
identifiers). The final dataset labels reflect this adjudicated 
outcome and serve as ground truth for all subsequent experiments.

\subsection{Dataset Statistics and Availability}

Table~\ref{tab:class_distribution} summarizes the overall class
distribution where genuine leaks account for 25.1\% of samples and
placeholders for 12.3\%, reflecting the natural
skew seen in real-world repositories.
Table~\ref{tab:language_class_distribution} further breaks down each
class by programming language. Notably, languages commonly used for
scripting and web development, such as Python, JavaScript, and Java exhibit, higher proportions of genuine leaks and placeholders, consistent with their prevalence in public repositories and tutorial code. We include this breakdown to demonstrate that credential ambiguity is not confined to any single language but arises across heterogeneous codebases, motivating a three-class formulation that generalizes across development contexts.

\section{Proposed Approach}
\label{sec:method}

\begin{figure*}[thbp]
    \centering
    \includegraphics[width=\linewidth]{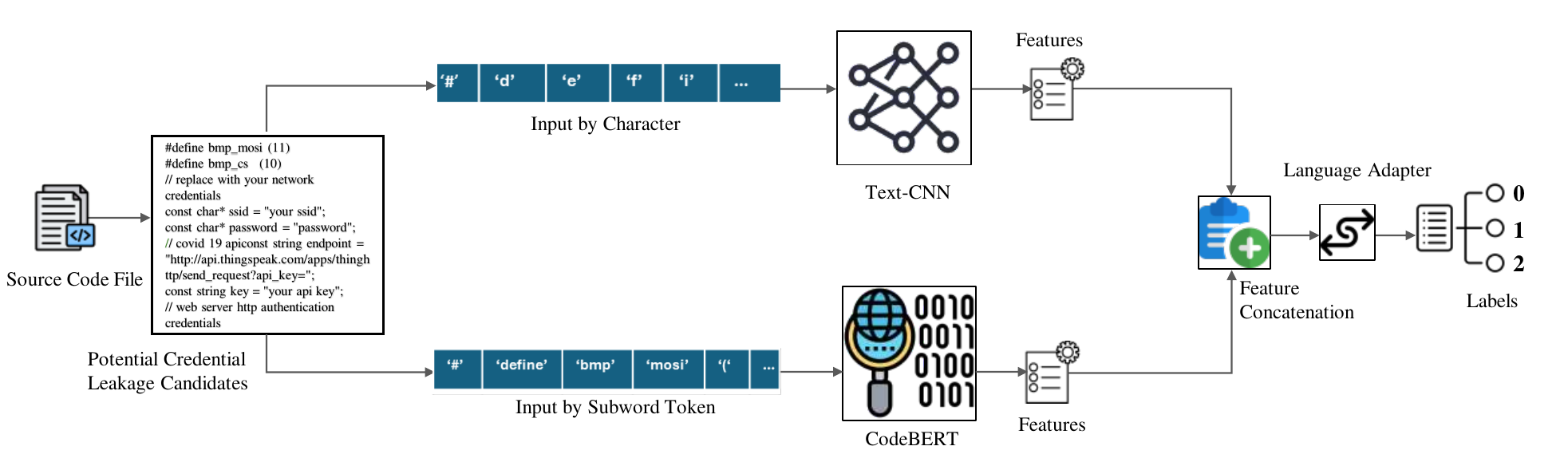}
    \caption{Overview of the proposed hybrid credential leakage detection model. A CodeBERT-based semantic encoder models surrounding code context to classify across all three classes: 0 (No Leak), 1 (Genuine Leak), and 2 (Placeholder), with a character-level encoder providing complementary syntactic coverage. The fused representation is refined by a language adapter and classifier to produce the final prediction.}
    \label{fig:overview}
\end{figure*}

Figure~\ref{fig:overview} provides an overview of the model architecture and information flow. A code snippet is processed through two parallel streams: (i) a CharCNN encoder analyzing the candidate string at the character level and (ii) a CodeBERT encoder modeling the surrounding code context, whose outputs are fused, refined by a language adapter, and passed to a classifier to predict one of three credential classes: No Leak, Genuine Leak, or Placeholder.

We use a dual-stream architecture that combines syntactic and semantic cues. In the syntactic stream, we analyze the candidate string at the character level by embedding individual characters and applying convolutional filters of different sizes. This design allows us to capture both short and long character sequences, such as delimiters, structured prefixes, and fixed-length tokens, as character-level convolutional models are known to learn these structural patterns effectively across diverse formats~\cite{zhang2015character}.

The second stream models the semantic code context surrounding the candidate string. We encode the enclosing code snippet using a pretrained CodeBERT model, which jointly represents source code and natural language comments~\cite{feng2020codebert}. This representation captures contextual cues such as variable naming, assignment structure, and usage patterns that indicate whether a value serves as a real credential, a placeholder, or a benign constant. Semantic context plays a critical role in disambiguating cases where syntactic appearance alone is insufficient, such as enum values or documentation examples discussed in Section~\ref{sec:dataset}.

We project the outputs of the two streams into a shared latent space and concatenate them to form a unified representation. A lightweight language adapter refines this representation using a feed-forward transformation with normalization and dropout. The adapter reduces language-specific variation while preserving signals relevant to credential semantics, supporting robustness across heterogeneous codebases without introducing language-specific modeling as a primary objective. A multi-layer perceptron classifier produces logits over the three credential classes. By combining these two streams within a three-class formulation, the proposed approach reduces alert noise and supports more effective triage in real-world software maintenance settings.

\section{Experimental Setup}
\label{sec:experiments}

This section describes the experimental protocol used to evaluate 
the proposed CLD approach. The baseline implementations and adaptations, evaluation metrics, and training configuration are described in the following subsections.

\subsection{Datasets and Splits}

To ensure consistency across methods and enable fair comparison, we construct a single stratified  80/10/10 split into training, validation, and test sets using a 
fixed random seed over the same dataset (as described in Section~\ref{sec:dataset}). We perform stratification jointly over class 
labels and programming languages to preserve both class balance 
and language diversity in each partition. We keep this data split 
fixed for all baselines and for our method.

To account for optimization variance, we train the proposed method 
using four different random initialization seeds (i.e., 42, 123, 456, and 789) 
while keeping the data partition unchanged. We report the mean and 
standard deviation across these runs. We use the validation set 
exclusively for hyperparameter tuning and early stopping, and we 
hold out the test set strictly for final evaluation.

\subsection{Baseline Methods}

We compare our approach against two state-of-the-art learning-based CLD systems. Since both prior works are originally designed for binary detection settings, we adapt them to support three-class snippet-level classification. To ensure fair comparison, we train all baselines with the same loss function, class weighting, and data split as our proposed method (described in Section~\ref{sec:training}). Any performance difference therefore reflects architectural choices rather than training strategies.

\textbf{PassFinder (Adapted).}
We adapt only the context model from PassFinder~\cite{feng2022icse}, 
which detects hard-coded passwords through character-level convolutional 
neural networks. The original architecture consists of a 6-layer 
character-level CNN with 256 filters per layer. To align PassFinder with 
our task formulation, we expand the output layer from binary (leak/no-leak) 
to a three-class classification. We maintain all other architectural 
components, including the original convolutional layer configuration 
(kernel sizes are 7, 3, 3, 3, 3, and 3), max pooling strategy, and fully 
connected layers with 50\% dropout.

\textbf{KEYSENTINEL (Adapted).}
We also compare against KEYSENTINEL~\cite{zhou2025sp}, the 
state-of-the-art CLD method. KEYSENTINEL extracts credential candidates 
using 910 regex patterns, applies a six-stage heuristic filtering 
pipeline (T1 through T6), and classifies candidates with a character-level 
TextCNN. Because snippets provide less context than complete files, we 
adjust KEYSENTINEL's filter thresholds. We lower T1's minimum length 
from 6 to 4 characters, reduce T3's entropy thresholds from 2.4/3.0 to 
2.0/2.5, and relax T4's word ratio threshold from 35\% to 50\%. We 
retrain the TextCNN classifier on candidate strings extracted from our 
labeled snippets using three-way classification. To produce snippet-level 
predictions, KEYSENTINEL aggregates its candidate-level outputs. The 
method labels a snippet as \textit{Genuine Leak} when any candidate 
exceeds confidence threshold $\tau=0.50$ or matches definitive patterns 
like AWS keys, JWTs, or Privacy-Enhanced Mail (PEM) blocks. Otherwise, KEYSENTINEL assigns 
\textit{Placeholder} when it detects placeholder markers ($\tau=0.60$), 
or \textit{No Leak} when neither condition holds.

\subsection{Training and Implementation Details}
\label{sec:training}

All models are implemented in PyTorch and trained on a single NVIDIA GPU. Our hybrid architecture combines a character-level CNN encoder ($d_c = 32$, 32 filters per kernel size in $\{3, 5, 7\}$, up to 800 characters) with CodeBERT-base~\cite{feng2020codebert} for semantic understanding (up to 256 tokens), with the first 6 layers frozen. Both encoders project to 384-dimensional representations before concatenation. The language adapter reduces the 768-dimensional fused space to 384 via a feed-forward network with ReLU, dropout, and layer normalization. The final classifier uses two fully connected layers (384 $\times$ 96 $\times$ 3) with dropout 0.2.

To address class imbalance, we employ focal loss~\cite{lin2017focal} ($\gamma = 2.0$) with inverse square-root class weighting. We use AdamW (weight decay 0.01, gradient clipping 1.0) with differentiated learning rates: $3 \times 10^{-5}$ for non-CodeBERT and $6 \times 10^{-6}$ for CodeBERT layers, with a linear decay scheduler to 10\%. Training runs for 15 epochs maximum with early stopping (patience = 5) on validation Matthews Correlation Coefficient (MCC) and batch size 32.
\section{Results and Analysis}
\label{sec:results}

We present a comprehensive evaluation of the proposed CLD approach organized around the five research questions stated in Section~\ref{sec:intro}. We report Seed 42 results as our primary evaluation unless otherwise stated; mean and standard deviation over four seeds (42, 123, 456, and 789) are reported to demonstrate robustness.

\subsection{RQ1: Dataset Quality and Reliability}
\begin{framed}
\noindent \textit{To what extent can a manually annotated dataset reliably capture 
credential occurrences across multiple programming languages?}
\end{framed}

Two annotators label each sample independently and achieve Cohen's 
$\kappa = 0.89$ before adjudication, which indicates near-perfect 
agreement~\cite{artstein2008inter}. This level of agreement is 
meaningful because the task is inherently difficult. Many credential-like 
strings look structurally similar, and correct labeling often requires 
reasoning about the surrounding code context rather than the string value 
alone. The three-class schema proves consistently interpretable across 
annotators, which confirms that security practitioners can reliably 
distinguish genuine leaks from placeholders.

All 10 languages contribute balanced test samples (88--97 instances 
each), and all 3 classes appear in every language. This confirms 
that placeholders are not specific to any single language 
but arise consistently across heterogeneous codebases.

\begin{tcolorbox}[colback=gray!10, colframe=gray!50, boxrule=0.5pt]
\textbf{RQ1 Summary:} The dataset reliably captures credential occurrences across 10 programming languages with near-perfect inter-annotator agreement ($\kappa = 0.89$), providing a balanced and representative foundation for evaluating the three-class credential detection.
\end{tcolorbox}

\subsection{RQ2: Three-Class Detection Effectiveness}

\begin{framed}
\noindent \textit{How effectively does the proposed three-class formulation detect 
genuine credential leaks while distinguishing them from placeholder or 
weak credentials?}
\end{framed}
\subsubsection{Per-Class Performance}
\begin{table}[t]
\centering
\caption{Per-class performance of the proposed method under Seed~42.}
\label{tab:perclass}
\begin{tabular}{lccc}
\toprule
\textbf{Class} & \textbf{Precision} & \textbf{Recall} & \textbf{F1} \\
\midrule
No Leak            & 0.96 & 0.93 & 0.95 \\
Genuine Leak       & 0.89 & 0.93 & 0.91 \\
Placeholder  & 0.78 & 0.83 & 0.81 \\
\bottomrule
\end{tabular}
\end{table}
In Table~\ref{tab:perclass} the model reports F1 of 0.95 for No Leak, 0.91 for Genuine Credential 
Leak, and 0.81 for Placeholder. The Genuine Leak recall of 0.93 is the most security-critical metric, where the model correctly identifies 222 out of 238 test cases, missing only 16 actual credentials. 
Placeholder detection shows higher recall (0.83) than precision (0.78), 
reflecting a conservative design choice where the model prefers to flag ambiguous 
strings as placeholders rather than risk misclassifying them as genuine 
threats.

Table~\ref{tab:robustness} confirms these results hold consistently across all four seeds. Per-class performance remains stable across initializations: No Leak F1 at 0.95±0.002, Genuine Leak F1 at 0.90±0.004, and Placeholder F1 at 0.80±0.008, confirming consistent learning regardless of initialization. The only moderate variance appears in placeholder precision (0.76±0.020), reflecting the inherent ambiguity of borderline cases that shift slightly across initializations.

\subsubsection{Error Analysis}
\begin{figure}[t]
    \centering
    \includegraphics[width=0.48\textwidth]{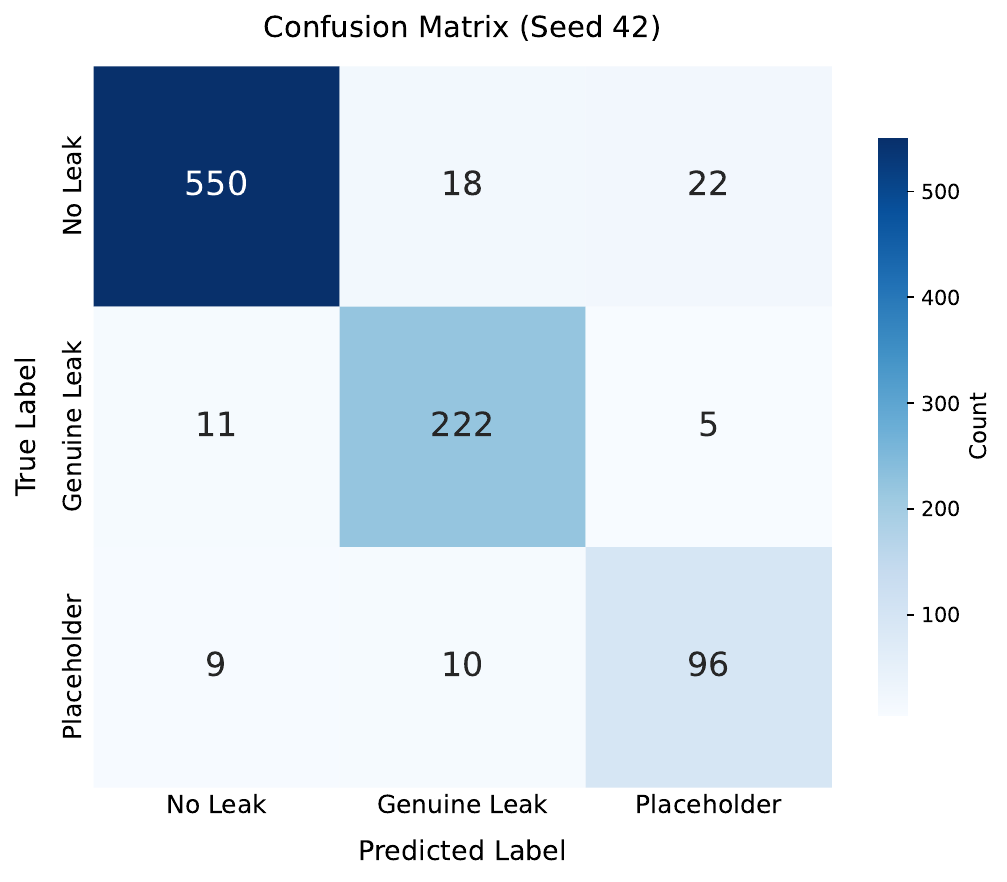}
    \caption{Confusion matrix of the proposed model under Seed~42. 
    Out of 943 test samples, 868 are correctly classified.}
    \label{fig:confusion}
\end{figure}

Figure~\ref{fig:confusion} shows the confusion matrix. Out of 943 test samples, the model misclassifies 75 cases (8.0\%). Three dominant error patterns emerge.
The most frequent errors involve No Leak misclassification: 18 samples are predicted as Genuine Leak and 22 as Placeholder. These arise when variable names carry strong credential signals 
(e.g., \texttt{password}, \texttt{api\_key}) even though no literal value is assigned. In such cases, the character-level encoder responds to the variable name rather than the absent value.

The most security-critical errors are five cases where the model predicts 
Placeholder instead of Genuine Leak. These involve weak, dictionary-like 
passwords (e.g., \texttt{"root"}, \texttt{"password"}) in low-specificity 
contexts that resemble test fixtures. JavaScript contributes two of these 
five cases, where test scaffolding and production code often share identical 
credential patterns.

Finally, 10 Placeholder samples are incorrectly flagged as Genuine Leaks. 
These involve test configurations where placeholder strings lack explicit 
syntactic markers. For instance, a C\# server configuration contains 
\texttt{Password = "youshallnotpass"}. This is a reference to the movie 
\textit{The Lord of the Rings} and is clearly a placeholder, not a real 
credential. However, the model flags it as genuine because the string 
lacks conventional placeholder markers such as underscores or capitalized 
instruction text like \texttt{YOUR\_PASSWORD}. Without explicit markers 
or surrounding comments indicating test usage, the model sometimes 
misinterprets creative placeholder text as potentially real credentials. 
Java and C\# together account for 4 of these 10 cases.

\begin{table}[t]
\centering
\caption{Per-class performance of the proposed method across four random seeds (Mean $\pm$ Std).}
\label{tab:robustness}
\small
\begin{tabular}{llc}
\toprule
\textbf{Class} & \textbf{Metric} & \textbf{Mean $\pm$ Std} \\
\midrule
No Leak          & Precision & 0.97 $\pm$ 0.004 \\
                 & Recall    & 0.93 $\pm$ 0.005 \\
                 & F1        & 0.95 $\pm$ 0.002 \\
                 & MCC       & 0.87 $\pm$ 0.006 \\
\midrule
Genuine Leak     & Precision & 0.88 $\pm$ 0.006 \\
                 & Recall    & 0.93 $\pm$ 0.009 \\
                 & F1        & 0.90 $\pm$ 0.004 \\
                 & MCC       & 0.87 $\pm$ 0.005 \\
\midrule
Placeholder & Precision & 0.76 $\pm$ 0.020 \\
                 & Recall    & 0.84 $\pm$ 0.007 \\
                 & F1        & 0.80 $\pm$ 0.008 \\
                 & MCC       & 0.77 $\pm$ 0.009 \\
\bottomrule
\end{tabular}
\end{table}

\noindent\textbf{Impact on Alert Triage.}
Binary credential detectors produce only two outcomes (alert vs.\ no alert) 
and therefore cannot differentiate between genuine leaks and placeholder artifacts; as a result, both are surfaced as alerts requiring manual inspection.
To quantify this maintenance impact, we simulate a binary alerting view in 
which any credential-like prediction (Genuine or Placeholder) is surfaced 
as an alert.
On the test set, this yields 373 alerts.
In contrast, our three-class model distinguishes placeholders from genuine 
leaks and produces 250 high-severity alerts (predicted Genuine only), 
corresponding to a 33.0\% reduction in high-severity alerts while 
maintaining 93\% recall for genuine leaks.

\begin{tcolorbox}[colback=gray!10, colframe=gray!50, boxrule=0.5pt]
\textbf{RQ2 Summary:} The three-class formulation effectively detects genuine leaks (recall: 0.93, F1: 0.91) while distinguishing placeholders (F1: 0.81), achieving overall model Macro-F1 of 0.90 and MCC of 0.86. This reduces high-severity alert volume by 33.0\% (373$\rightarrow$250) without sacrificing security coverage.
\end{tcolorbox}

\subsection{RQ3: Cross-Language Generalization}

\begin{framed}
\noindent \textit{How well does the approach generalize across programming languages, 
and what role does the language adapter play?}
\end{framed}

\subsubsection{Leave-One-Language-Out Cross-Validation} 
To assess cross-language generalization, we conduct Leave-One-Language-Out 
(LOLO) cross-validation, where we train on nine languages and evaluate on 
the held-out language. This protocol tests whether the model can generalize 
to entirely unseen programming languages without any language-specific 
training data.
 
Table~\ref{tab:cross_language_results} shows LOLO results. The model achieves 
an average F1 of 0.83 and MCC of 0.79 across all 10 languages, demonstrating 
robust generalization to completely unseen languages. Performance remains 
strong even without language-specific training data, with 9 languages achieving $F_1 \geq 0.80$. C, Java, and Python show particularly strong LOLO performance ($F_1 \geq 0.85$), while Go and TypeScript show slightly lower but still reasonable performance ($F_1$: 0.76, 0.80, respectively). The consistent performance 
across diverse language families confirms that the model learns language-invariant 
credential patterns rather than overfitting to language-specific syntax.
\begin{table}[t]
\centering
\caption{LOLO cross-validation results. Model trained on nine languages, 
evaluated on an unseen target language.}
\label{tab:cross_language_results}
\small
\begin{tabular}{lcccc}
\toprule
\textbf{Language} & \textbf{Prec} & \textbf{Rec} & \textbf{F1} & \textbf{MCC} \\
\midrule
C            & 0.87 & 0.91 & 0.89 & 0.85 \\
C\#          & 0.79 & 0.89 & 0.82 & 0.80 \\
C++          & 0.80 & 0.88 & 0.84 & 0.80 \\
Go           & 0.75 & 0.78 & 0.76 & 0.72 \\
Java         & 0.89 & 0.88 & 0.89 & 0.84 \\
JavaScript   & 0.83 & 0.85 & 0.84 & 0.78 \\
PHP          & 0.82 & 0.86 & 0.83 & 0.78 \\
Python       & 0.86 & 0.85 & 0.85 & 0.78 \\
Ruby         & 0.82 & 0.86 & 0.83 & 0.78 \\
TypeScript   & 0.76 & 0.85 & 0.80 & 0.74 \\
\midrule
\textbf{Average} & \textbf{0.81} & \textbf{0.86} & \textbf{0.83} & \textbf{0.79} \\
\bottomrule
\end{tabular}
\end{table} 
\subsubsection{Impact of Language Adapter}

Having established a strong baseline cross-language generalization, we now examine whether the language adapter further improves performance. Table~\ref{tab:language_adapter_comparison} compares the full model with the no-adapter variant across all 10 languages.

\begin{table}[t]
\centering
\caption{Per-language MCC: Full Model vs.\ No Adapter (Seed~42).
Blue/red $\Delta$MCC indicates gain/loss from the adapter.}
\label{tab:language_adapter_comparison}
\small
\begin{threeparttable}
\begin{tabular}{lccc}
\toprule
\textbf{Language} & \textbf{MCC$_{\text{Full}}$} & \textbf{MCC$_{\text{NoAdp}}$} & \textbf{$\Delta$MCC} \\
\midrule
C          & \textbf{82.9} & 79.6          & {\color{blue}$+3.3$} \\
C\#        & \textbf{91.6} & 89.2          & {\color{blue}$+2.4$} \\
C++        & \textbf{79.8} & 76.5          & {\color{blue}$+3.3$} \\
Go         & \textbf{75.5} & 75.5          & $0.0$ \\
Java       & \textbf{85.5} & 82.2          & {\color{blue}$+3.3$} \\
JavaScript & \textbf{88.5} & 80.8          & {\color{blue}$+7.7$} \\
PHP        & 83.0          & \textbf{87.6} & {\color{red}$-4.6$} \\
Python     & 77.5          & \textbf{77.7} & {\color{red}$-0.2$} \\
Ruby       & \textbf{92.6} & 90.9          & {\color{blue}$+1.7$} \\
TypeScript & \textbf{82.0} & 80.1          & {\color{blue}$+1.9$} \\
\bottomrule
\end{tabular}
\begin{tablenotes}
\footnotesize
\item All values in \%. $\Delta$MCC = MCC$_{\text{Full}}$ minus MCC$_{\text{NoAdp}}$.
\end{tablenotes}
\end{threeparttable}
\end{table}

The language adapter provides consistent benefits across 8 of 10 languages. 
JavaScript achieves the largest gain, i.e., 7.7 percentage points (pp) increase in MCC, suggesting the adapter effectively handles its loosely-typed variable patterns, dynamic string construction, and diverse credential assignment idioms. For C, Java, and C++, all of them show moderate improvements (+3.3 pp), indicating that the adapter captures useful syntactic signals even in statically-typed languages where credential patterns are more rigidly structured. C\# (+2.4 pp), TypeScript (+1.9 pp), and Ruby (+1.7 pp) show smaller but consistent gains, while Go exhibits a neutral effect (0.0 pp).

Two languages show degradation: PHP ($-$4.6 pp) and Python ($-$0.2 pp). PHP's decline suggests that language-specific conditioning may introduce noise when the adapter's learned representation conflicts with PHP's weakly-typed string interpolation and mixed quoting conventions. Python's drop of 0.2 pp is negligible and falls within the expected single-seed variance, effectively constituting a neutral result.

Overall, the adapter improves or preserves performance for 9 of 10 languages, with the single meaningful degradation isolated to PHP. This pattern indicates that language-specific conditioning is broadly beneficial 
across diverse language families, particularly for languages with heterogeneous string handling and dynamic typing with PHP standing as a notable exception warranting further investigation.

\begin{tcolorbox}[colback=gray!10, colframe=gray!50, boxrule=0.5pt]
\textbf{RQ3 Summary:} The model demonstrates robust cross-language generalization 
through LOLO evaluation (average $F_1$: 0.83, MCC: 0.79), with 9 languages 
achieving $F_1 \geq 0.80$ even when completely unseen during training. The language 
adapter provides consistent gains across 8 of 10 languages, with JavaScript 
benefiting the most (+7.7 pp MCC). PHP is the primary exception ($-$4.6 pp), 
while Python shows a negligible neutral effect ($-$0.2 pp), indicating broadly 
beneficial but not universal adaptation.
\end{tcolorbox}

\subsection{RQ4: Architectural Component Contributions}
\begin{framed}
\noindent \textit{What is the contribution of each architectural component to overall detection performance?}
\end{framed}

Table~\ref{tab:ablation} presents a comprehensive ablation study in which we systematically isolate the contribution of each component. We evaluate six variants to understand both individual and combined effects of the character-level encoder, semantic encoder, and language adapter.

\begin{table}[t]
\centering
\caption{Ablation study showing component contributions (Seed~42). 
Full Model uses CharCNN+CodeBERT+Adapter; other variants isolate 
individual components and combinations.}
\label{tab:ablation}
\small
\begin{tabular}{lcccc}
\toprule
\textbf{Variant} & \textbf{MCC} & \textbf{F1} & \textbf{Prec} & \textbf{Rec} \\
\midrule
Full Model          & 0.86 & 0.90 & 0.89 & 0.91 \\
\midrule
CharCNN Only        & 0.62 & 0.70 & 0.68 & 0.72 \\
CharCNN + Adapter   & 0.57 & 0.67 & 0.66 & 0.70 \\
\midrule
CodeBERT Only       & 0.85 & 0.88 & 0.86 & 0.90 \\
CodeBERT + Adapter  & 0.85 & 0.88 & 0.87 & 0.90 \\
\midrule
No Adapter          & 0.85 & 0.88 & 0.87 & 0.90 \\
\bottomrule
\end{tabular}
\end{table}

The semantic code encoder (CodeBERT) is the dominant component. Removing 
it causes performance to drop substantially (MCC: 0.86 $\rightarrow$ 0.57, 
F1: 0.90 $\rightarrow$ 0.67), demonstrating that understanding surrounding code context is essential for distinguishing placeholders from genuine credentials. The CodeBERT-only variant achieves an MCC of 0.85, and the full model improves upon this by 1.0 pp (0.85 $\rightarrow$ 0.86), indicating that character-level features provide a complementary signal beyond what semantic context alone captures.

We retain the character-level encoder for two reasons. First, character patterns provide targeted coverage for highly structured credentials---such as AWS access keys (fixed 20-character alphanumeric format), JWTs (three dot-delimited Base64 segments), and PEM-encoded private keys (structured header/footer delimiters)---that follow rigid surface formats, which CodeBERT's subword tokenization may not preserve faithfully. Second, the computational overhead is negligible (32 filters vs. CodeBERT's 768-dimensional embeddings), making retention a low-cost robustness measure 
for deployment scenarios where such structured tokens appear outside the training distribution.

The language adapter yields a 1.0 pp aggregate MCC gain (0.85 
$\rightarrow$ 0.86), with the largest benefit for JavaScript 
(+7.7 pp) and consistent moderate gains for C, Java, and C++ 
(+3.3 pp each). PHP is the primary exception ($-$4.6 pp), and 
Python shows a negligible neutral effect ($-$0.2 pp). 
Practitioners working primarily in PHP may prefer the 
no-adapter variant.

\begin{tcolorbox}[colback=gray!10, colframe=gray!50, boxrule=0.5pt]
\textbf{RQ4 Summary:} The semantic code encoder is the critical component (MCC reduces by 29.0 pp when it is removed). The three-class formulation is primarily enabled by CodeBERT's contextual understanding, while character-level features provide targeted complementary coverage for structurally rigid credential formats such as AWS keys and JWTs that undergo subword tokenization may be fragmented. The full model improves upon CodeBERT-only by 1.0 pp MCC at negligible computational cost. The language adapter contributes further 1.0 pp aggregate gain, improving performance across 8 of 10 languages, with mixed effects for PHP and Python as detailed in RQ3.
\end{tcolorbox}

\subsection{RQ5: Comparison with State-of-the-Art Systems}

\begin{framed}
\noindent \textit{How does the proposed approach compare with state-of-the-art CLD systems under a unified three-class evaluation?}
\end{framed}

\begin{figure}[t]
    \centering
    \includegraphics[width=0.48\textwidth]{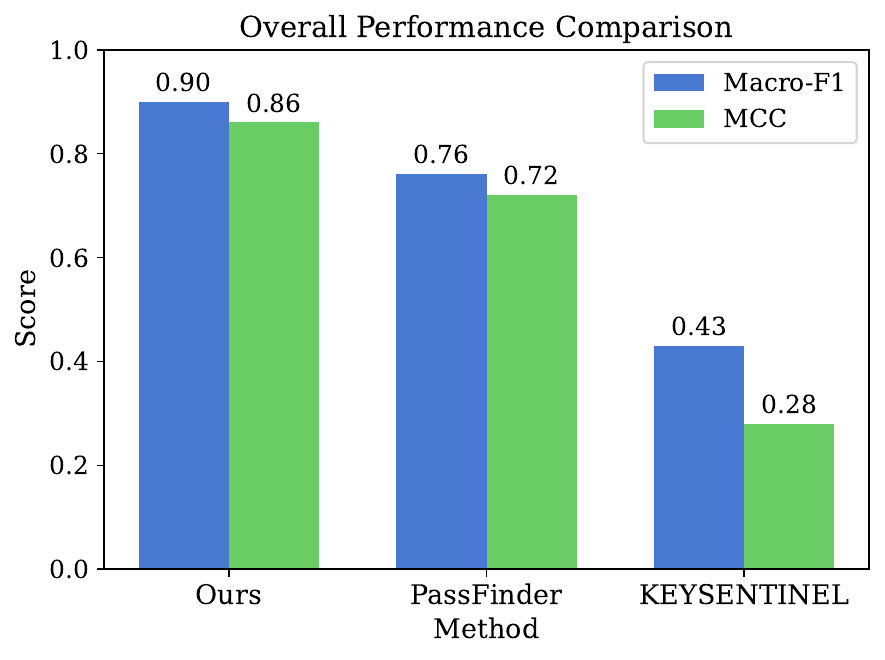}
    \caption{Overall performance comparison across methods. Our approach 
    substantially outperforms both baselines in Macro-F1 and MCC.}
    \label{fig:overall_comparison}
\end{figure}

Figure~\ref{fig:overall_comparison} and Table~\ref{tab:perclass_comparison} 
present comprehensive comparisons between our approach and two state-of-the-art 
learning-based credential detection systems adapted to the three-class setting.

\begin{table}[t]
\centering
\caption{Per-class performance comparison across three methods (Seed~42).
PF = PassFinder; KS = KEYSENTINEL.}
\label{tab:perclass_comparison}
\small
\setlength{\tabcolsep}{4pt}
\begin{tabular}{llccc}
\toprule
\textbf{Class} & \textbf{Metric} & \textbf{Ours} & \textbf{PF} & \textbf{KS} \\
\midrule
No Leak          & Precision & \textbf{0.96} & 0.96 & 0.67 \\
                 & Recall    & 0.93 & 0.93 & \textbf{0.94} \\
                 & F1        & \textbf{0.95} & 0.93 & 0.78 \\
\midrule
Genuine Leak     & Precision & 0.89 & 0.84 & \textbf{0.94} \\
                 & Recall    & \textbf{0.93} & 0.79 & 0.30 \\
                 & F1        & \textbf{0.91} & 0.81 & 0.46 \\
\midrule
Placeholder      & Precision & \textbf{0.78} & 0.50 & 0.09 \\
                 & Recall    & \textbf{0.83} & 0.58 & 0.03 \\
                 & F1        & \textbf{0.81} & 0.54 & 0.04 \\
\bottomrule
\end{tabular}
\end{table}
\subsubsection{Comparison with PassFinder}

PassFinder relies primarily on character-level patterns with limited semantic 
context modeling, restricting its ability to resolve ambiguous credential-like 
strings. Our approach improves overall Macro-F1 and MCC by 14.0 pp each (0.76 → 0.90 and 0.72 → 0.86, respectively). The most significant improvement occurs in the placeholder class, where our F1 score of 0.81 exceeds PassFinder's 0.54 by 27.0 pp. This supports our hypothesis that combining semantic code context with character-level features improves disambiguation of placeholders that resemble genuine secrets at the surface level but differ in their contextual usage.

For genuine credential detection, our approach also improves over PassFinder (F1: 0.91 vs. 0.81), driven by higher recall (0.93 vs. 0.79). This improvement stems from semantic understanding of credential usage patterns, enabling 
detection of low-entropy human-chosen passwords that lack distinctive character patterns but appear in security-relevant contexts.

\subsubsection{Comparison with KEYSENTINEL}
KEYSENTINEL was originally designed for file-level detection. Our evaluation adapts it to snippet-level classification to assess architectural trade-offs under a unified setting. It achieves Macro-F1 of 0.43 and MCC of 0.28 in our snippet-level evaluation. The results reveal an important finding about heuristic filtering approaches. The genuine leak recall of 0.30 indicates that extensive filtering designed to reduce false positives in file-level detection inadvertently discards 70\% of real credentials when applied to limited-context snippets. This demonstrates a fundamental trade-off in context-discarding paradigms: aggressive filtering that works well for cross-platform file-level detection (KEYSENTINEL's core contribution) becomes overly restrictive when context is limited.

Placeholder detection fails almost completely (F1: 0.04, recall: 0.03) because 
low-entropy placeholder strings (e.g., \texttt{"YOUR\_API\_KEY"}, 
\texttt{"dummy\_token"}) fail T3 entropy filters, while word-heavy placeholders 
fail T4 semantic filters. Both filter stages were designed to remove 
non-credentials, but cannot distinguish intentional placeholders from weak 
credentials without surrounding code context.

These results highlight complementary strengths: KEYSENTINEL's cross-platform filtering excels in file-level detection, where broader context supports candidate retention, while our context-preserving semantic approach better handles snippet-level scenarios common in code review and pull request workflows.

\subsubsection{Three-Class Detection Value}

The substantial improvements in placeholder identification (F1: 0.81 vs. baseline range 0.04-0.54) demonstrate clear practical value of explicit three-class formulation. In deployment scenarios, flagging placeholders as a separate category enables developers to triage alerts more effectively, reducing alert fatigue without sacrificing security coverage. The high recall for genuine leaks (0.93) ensures that critical security threats are rarely missed despite the additional class complexity.

\begin{tcolorbox}[colback=gray!10, colframe=gray!50, boxrule=0.5pt]
\textbf{RQ5 Summary:} The proposed approach substantially outperforms adapted baselines. Compared to PassFinder, we improve Macro-F1 by 14.0 pp and placeholder detection by 27.0 pp. KEYSENTINEL's low snippet-level performance (30\% genuine leak recall) reveals that aggressive heuristic filtering designed for file-level detection inadvertently discards real credentials when context is limited, demonstrating complementary strengths: cross-platform filtering for files versus semantic context preservation for snippets.
\end{tcolorbox}

\section{Threats to Validity}\label{sec:threats}

\textbf{Construct Validity.} Labeling credential data involves judgment, particularly for borderline placeholder cases. We mitigate this using a three-stage annotation workflow with near-perfect inter-annotator agreement (Cohen's $\kappa$ = 0.89), where genuinely ambiguous cases are conservatively assigned to the Genuine Leak class to minimize underestimation of security impact. The binary alert simulation in RQ2 surfaces both Genuine and Placeholder predictions as undifferentiated alerts, reflecting how a two-class system behaves without reinterpreting placeholder labels.

\textbf{Internal Validity.} All methods use identical data splits, training procedures, and evaluation protocols. Results are reported across four random seeds to account for initialization variance. Hyperparameters were tuned exclusively on the validation set with a strict test set holdout.

\textbf{External Validity.} Evaluation spans 10 programming 
languages and public repositories. Results may not fully 
generalize to proprietary or highly obfuscated code. The 
reported 33.0\% alert reduction reflects test-set class 
proportions; actual reduction in deployment will vary with 
repository-level placeholder rates. However, Leave-One-Language-Out 
validation (average F1: 0.83; all languages F1 $\geq$ 0.76) 
indicates that the model captures language-invariant credential patterns. Snippet-level evaluation (up to 13 lines) reflects realistic review settings but excludes repository-level context.

\textbf{Baseline Adaptation.} PassFinder and KEYSENTINEL were adapted for three-class, snippet-level classification while preserving core architectures. Both were trained under identical settings for fairness. Lower snippet-level recall (e.g., 30\% for KEYSENTINEL) likely reflects adaptation to a finer-grained task rather than inherent method weakness, highlighting trade-offs in context-dependent approaches.

\section{Conclusion and Future Work}\label{sec:conclusion}

This paper introduces a three-class formulation for CLD that 
explicitly separates genuine credential leaks, placeholders, 
and benign code. Evaluation on 9,426 manually annotated snippets 
across 10 programming languages demonstrates strong performance 
(Macro-F1: 0.90, MCC: 0.86, Genuine Leak recall: 0.93) with robust 
cross-language generalization (LOLO average F1: 0.83), reducing 
high-severity alerts by 33.0\% without 
sacrificing security coverage. Compared to prior character-level methods, placeholder detection improves from 54\% to 81\% F1-score. Ablation studies confirm semantic code understanding as the critical 
component, with character-level features providing complementary 
coverage and the language adapter, improving 8 of 10 programming languages. 

Future work includes, but is not limited to: (i) incorporating repository-level context (e.g., dependency structure and version history) to improve disambiguation, (ii) extending detection beyond source code to non-code artifacts where credentials commonly appear (e.g., configuration files and documentation), and (iii) investigating LLM-based augmentation to better handle hard borderline cases, e.g., via active learning with LLM-assisted annotation and targeted synthesis of ambiguous placeholders. In addition, we plan to conduct user studies to assess whether multi-class triage reduces alert fatigue in practice, and to explore lightweight deployment variants suitable for continuous integration.

\section*{Data Availability}
The artifacts, including source code and sanitized datasets, are publicly available at Zenodo (\href{https://doi.org/10.5281/zenodo.18881159}{doi:10.5281/zenodo.18881159}). Due to  security and privacy concerns, only the Class 0 (No Leak) samples  are released publicly, as Classes 1 and 2 contain credential-like strings that could expose sensitive patterns if made openly available. %

\section*{Acknowledgment}
This work was supported in part by the National Science Foundation under Grant No. 2347249. The authors would also like to thank Poorva Shetye for her assistance with the initial data collection.

\bibliographystyle{ieeetr}
\bibliography{refs}

\end{document}